\documentclass[slac_one]{revtex4}
\pdfoutput=1
\usepackage{graphicx}
\usepackage{fancyhdr}
\def\xsec{cm$^{-2}$s$^{-1}$}

\begin{document}

\title{$W$ Mass as a Calibration of the Jet Energy Scale at ATLAS} 

%

\author{Daniel Goldin, on behalf of the ATLAS Collaboration}
\affiliation{Department of Physics, Southern Methodist University, 
Dallas, Texas 75275, USA}

\begin{abstract}
Top-antitop pairs will be copiously produced at the LHC, at a rate of
roughly one per second at an instantaneous luminosity of $10^{33}$
\xsec. These events have low background and produce large numbers of
jets via the hadronic decay of the W's which may be used to calibrate
the jet energy scale and resolution with experimental data and
simulations. 
\end{abstract}

\maketitle

\thispagestyle{fancy}

\section{JET RECONSTRUCTION AT ATLAS} 
The ATLAS calorimeter \cite{atlas} is non-compensating. Its
calibration may be performed either {\em globally} or {\em
locally}. Global calibration is a top-down technique where hadronic
objects from specific physics processes are reconstructed relative to
a fixed (global) electromagnetic scale. Cells may be combined along
the $\eta$-$\phi$ direction to form cell towers, out of which {\em
tower jets} may be reconstructed after applying cell and calorimeter
layer weights based on the deposited jet energy density. The local
calibration method is a bottom-up approach which starts with the
response of cells and cell clusters, to which sets of weights based on
the ratio of the deposited energy to the signal reconstructed on the
electromagnetic energy scale, as determined with simulations, are
applied. From such calibrated topological clusters we reconstruct {\em
topological jets}.

In what follows we present a simulation study designed to assess the
ATLAS calorimeter performance through the jet energy scale (JES)
determination from the top-antitop pair simulation.
%
The work on the universal JES definition is still in progress. For the
study presented here we use two simple formulas:
\begin{equation}
  E^{true}_{jet} = K_{abs}(E,\eta, \sigma_j,L) E^{meas}_{jet}\,\,\,\mathrm{and}\,\,\,
  E_{part} = K_{part}(E,\Delta R, L) E^{true}_{jet} 
\end{equation}
The first of the two equations relates the energy $E^{true}_{jet}$ of
the true (particle) jet to that of the measured (reconstructed) jet
$E^{meas}_{jet}$. The scale factor $K_{abs}$ depends on energy,
pseudorapidity ($\eta$), jet resolution ($\sigma_j$) and instantaneous
luminosity ($L$). The second equation provides a transition from the
true jet to the quark that gave rise to the jet. The true jet-parton
scale factor $K_{part}$ depends on the parton energy, jet cone size
$\Delta R \equiv \Delta \eta \times \Delta \phi$ and the pileup
resulting from additional proton collisions in a bunch crossing, which
is proportional to the instantaneous luminosity $L$.

\section{JET ENERGY SCALE FROM THE TOP-ANTITOP SIGNAL}
\begin{figure}[!htp]
\vspace{-0.8cm}
\centering
\includegraphics[height=3.5cm,width=0.3\textwidth]{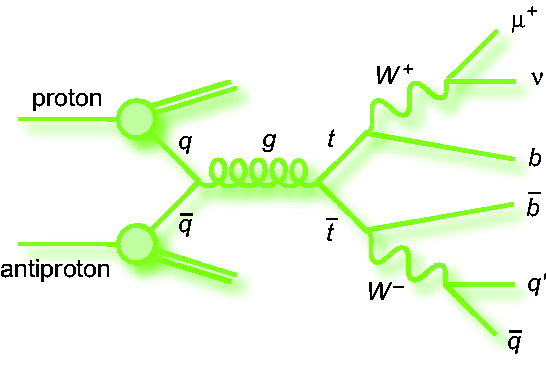}
\caption{Diagrams corresponding to the physics channels considered in this
work: top-antitop process with subsequent semileptonic decays.}
\label{fig:feyn}
\end{figure}
\begin{figure}
\centering
\includegraphics[height=5cm,width=0.4\textwidth]{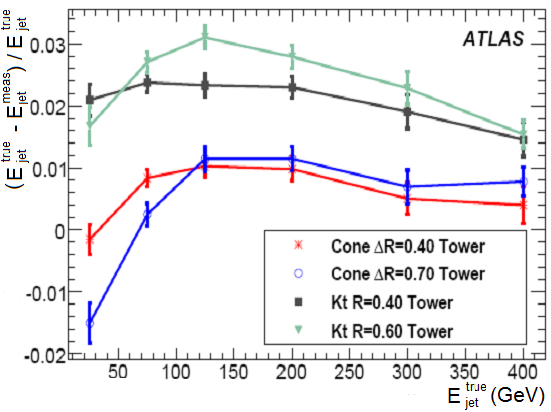}
\includegraphics[height=5cm,width=0.4\textwidth]{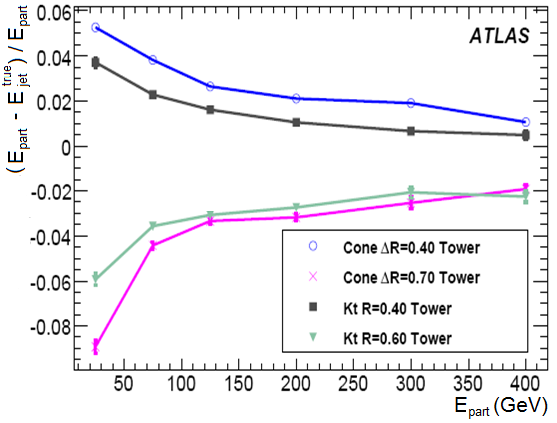}
\caption{Relative deviation of the reconstructed jet energy from the true jet 
energy as a function of the true jet energy (left) and the deviation
of the true jet energy from the parton energy as a function of the
parton energy (right) for various jet algorithms. Seeded cone jets
with $\Delta R = 0.4$ and 0.7 and $k_t$ jets with $D = 0.4$ and 0.6
were applied. The figures are used to extract $K_{abs}$ and
$K_{part}$.}
\label{fig:Kttbar}
\end{figure}

The top-antitop events considered here (Fig. \ref{fig:feyn}, left)
have been generated using MC@NLO \cite{mcatnlo} and simulated with
GEANT4 \cite{geant}. Four types of reconstructed events with various
jet settings were compared: the seeded cone jet algorithm
\cite{cone-alg} with parameter $\Delta R$ set at 0.4 and 0.7, and the
$k_t$ algorithm \cite{kt-alg} with $D = 0.4$\footnote{In literature
the parameter $D$ of the $k_t$ algorithm is also commonly referred to
as $R$.} and 0.6. Among the various cuts applied, events with at least
4 jets with $p_T > 40$ GeV/$c$ were chosen, and pairs of
non-$b$-tagged jets are selected as the $W$ jet candidates. Partons in
this case are quarks from the $W$ decay. $K_{abs}$ and $K_{part}$ may
be extracted from Fig. \ref{fig:Kttbar}. Using $(E^{true}_{jet} -
E^{meas}_{jet})/E^{true}_{jet} = 1 - 1/K_{abs}$ and $(E_{part} -
E^{true}_{jet})/E_{part} = 1 - 1/K_{part}$, for cone jets with $\Delta
R = 0.4$ we find $K_{abs}=1.032$ and $K_{part}=1.006$.

\begin{figure}[!htp]
\centering
\includegraphics[height=4.8cm,width=\textwidth]{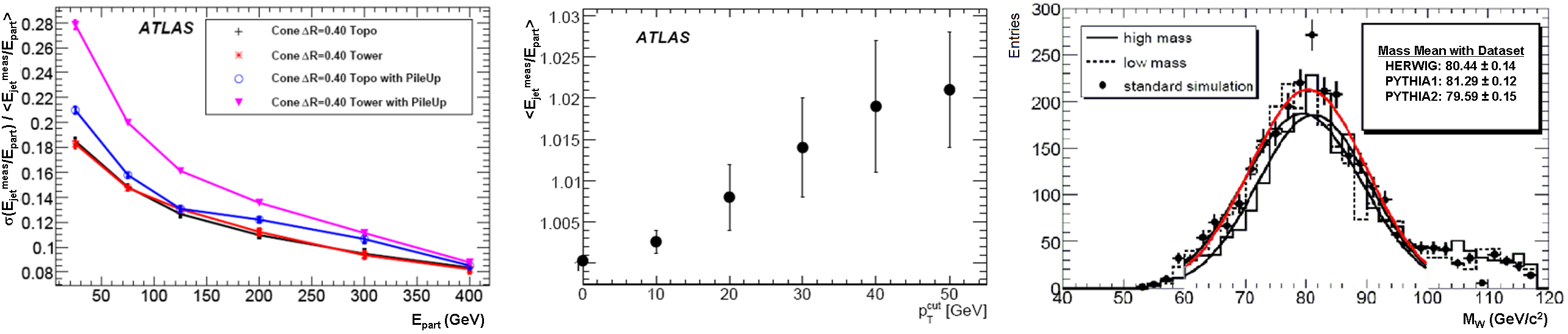}
\caption{Systematic effects in determining the JES from light jets in 
the top-antitop production. Left: effect of the pileup on the jet
resolution. Center: effect of the $p_T$ cut on the JES. Right: Results
of the fit to the $W$ mass from light jets for different gluon
radiation settings.} 
\label{pic:jes_sys}
\end{figure}

From Fig. \ref{fig:Kttbar} it is clear that the performance of the
reconstructed jets relative to the true jets and true jets relative to
the partons for the energy range shown is best for the seeded cone jet
algorithm with $\Delta R = 0.4$ and the $k_t$ algorithm with $D =
0.4$. We select the cone algorithm with $\Delta R = 0.4$ as the
default and use it on jets reconstructed from towers and from
topological clusters (Fig. \ref{pic:jes_sys}, left). We observe that,
without pileup effects included, the resolution does not change with
the reconstruction method and note that it varies between 8 and 19\%
for the energy range shown. In the same figure we also consider the
effects of pileup, a consequence of the increased luminosity giving
rise to the proportional number of multiple interactions. For a
luminosity of $10^{33}$ \xsec~the expected average number of these
pileup events is 2.3. If pileup is included, topological clustering,
which results in better noise suppression, yields better resolution,
especially in the lower energy range. For example, for jet energies
around 50 GeV, average energy resolution is decreased by only 2.5\%
relative to the pileup-free events, whereas the tower method yields
10\% worse resolution. Cone algorithm with $\Delta R = 0.4$
reconstructed from the topological clusters is thus the best choice
for the jets in the top-antitop process.

An applied $p_T$ cut on the measured jets biases the available
momentum phase space toward higher $p_T$ values (hence, higher energy
values) resulting in higher values of the jet energy scale
(Fig. \ref{pic:jes_sys}, center). The error bars correspond to
uncertainties in $p_T$ if the resolution is varied by 20\% (roughly
the worst case scenario according to the results above). The 40 GeV/$c$
$p_T$ cut used throughout this analysis thus results in a 2\%
uncertainty. For a 40 GeV/$c$ $p_T$ cut the bias on the $W$ mass was found
to be about $1.5 \pm 0.6$ GeV/$c^2$.

In order to study the influence of initial and final-state radiation,
three datasets with different showering settings were used: one set
with MC@NLO + HERWIG \cite{mcatnlo}\cite{herwig} and two sets with
AcerMC + PYTHIA \cite{acermc}\cite{pythia} (Fig. \ref{pic:jes_sys},
right). The mean mass from the three measurements is slightly higher
than the PDG value due to the $p_T$ cut bias. The spread between the
maximum and the minimum values is about 1.7 GeV/$c^2$.

\section{CONCLUSIONS}
%
In this work we have determined the values for $K_{abs}$, the measured
jet-to-true jet scale, and $K_{part}$, the true jet-parton energy
scale, for the top-antitop channel. Events were reconstructed using
local calibration. The value of $K_{abs}$ was found to be 1.032 for
the top-antitop signal. We also considered jet resolution, shown to
vary between 8 and 19\%, and examined the effects of the $p_T$ cut, 
gluon radiation settings and pileup on the jet energy scale for the
t-tbar signal.
\vspace{-0.7cm}

\begin{acknowledgments}
\vspace{-0.3cm}
The author gratefully acknowledges James Proudfoot, Jerome
Schwindling, David Miller, Ryszard Stroynowski, Jingbo Ye, Peter Loch
and Bruce Mellado for their help in preparation of these
proceedings. 
\vspace{-0.5cm}
\end{acknowledgments}

\end{document}